\documentclass[preprint,showpacs]{revtex4}
\usepackage{epsfig}
\newcommand{\ben}{\begin{eqnarray}}
\newcommand{\een}{\end{eqnarray}}

\newcommand{\bef}{\begin{figure}[h!bt]\centering}
\newcommand{\eef}{\end{figure}}
\newcommand{\bet}{\begin{table}[hbt]\centering}
\newcommand{\eet}{\end{table}}

\begin{document}
\title{Phase diagrams of ${\rm Ba(Fe}_{1-x}{\rm TM}_x)_2{\rm As}_2$  (TM =
Rh, Pd) single crystals}
\author{N. Ni}
\affiliation{Ames Laboratory and
Department of Physics and Astronomy, Iowa State University, Ames,
IA 50011, USA}

\author{A. Thaler}
\affiliation{Ames Laboratory and Department of Physics and
Astronomy, Iowa State University, Ames, IA 50011, USA}

\author{A. Kracher}
\affiliation{Ames Laboratory and
Department of Physics and Astronomy, Iowa State University, Ames,
IA 50011, USA}

\author{J. Q. Yan}
\affiliation{Ames Laboratory and
Department of Physics and Astronomy, Iowa State University, Ames,
IA 50011, USA}

\author{S. L. Bud'ko}
\affiliation{Ames Laboratory and Department of Physics and
Astronomy, Iowa State University, Ames, IA 50011, USA}

\author{P. C. Canfield}
\affiliation{Ames Laboratory and Department of Physics and
Astronomy, Iowa State University, Ames, IA 50011, USA}

\begin{abstract}
Single crystalline ${\rm Ba(Fe}_{1-x}{\rm TM}_x)_2{\rm As}_2$  (TM =
Rh, Pd) series have been grown and characterized by structural, thermodynamic and transport
measurements. These measurements show that the
structural/magnetic phase transitions, found in pure ${\rm BaFe}_2{\rm As}_2$ at 134 K, are suppressed monotonically by the doping and that
superconductivity can be stablized over a dome-like region. Temperature-composition ($T-x$) phase diagrams based on electrical transport and magnetization measurements
are constructed and compared to those of the ${\rm Ba(Fe}_{1-x}{\rm TM}_x)_2{\rm As}_2$  (TM =
Co, Ni) series. Despite the generic difference between $3d$ and $4d$ shells and the specific, conspicuous differences in the changes to the
unit cell parameters, the effects of Rh doping are exceptionally similar to the effects of Co doping and the
effects of Pd doping are exceptionally similar to the effects of Ni
doping. These data show that whereas the structural / antiferromagnetic phase transition temperatures can be parameterized
by $x$ and the superconducting transition temperature can be parameterized by some combination of $x$ and $e$, the number of extra electrons
associated with the TM doping, the transition temperatures of $3d-$ and $4d-$ doped ${\rm BaFe}_2{\rm As}_2$ can not be simply parameterized by the changes in the unit
cell dimensions or their ratios.

\end{abstract}
\date{\today}
\maketitle

\section{introduction}

The discovery of superconductivity in F-doped LaFeAsO \cite{yy} and K-doped ${\rm BaFe}_2{\rm As}_2$ \cite{Rotter}
compounds in the first half of 2008 has led
to extensive experimental interest; $T_c$ has risen as high as 56 K for F doped RFeAsO systems \cite{zhaozx} and as high as 38 K in
K and Na doped ${\rm (AE)Fe}_2{\rm As}_2$ systems (AE: Ba, Sr, Ca) \cite{Rotter}. Soon after, superconductivity was also
found in Co and Ni doped ${\rm (AE)Fe}_2{\rm As}_2$ \cite{A1,Nidope} and RFeAsO \cite{A2}. Recently, superconductivity was
also found in $4d$ and $5d$ transition metal electron doped ${\rm SrFe}_2{\rm As}_2$ \cite{srru,srpd,srrh,srir}. Although
electron doped ${\rm (AE)Fe}_2{\rm As}_2$ systems have
lower $T_c$ values ($\sim$24 K) \cite{NiCo,chu, nmr, wen}, intensive studies have been made on them because
doping is more homogeneous in these systems and the single crystals can be easily grown and reproduced.
For example, several studies of ${\rm Ba(Fe}_{1-x}{\rm Co}_x)_2{\rm As}_2$ system have resulted in remarkably similar data and
conclusions \cite{NiCo,chu, nmr, wen}.
In order to compare the effects of $3d$ and $4d$ electron doping in ${\rm BaFe}_2{\rm As}_2$, and thus try to
understand the conditions for the appearance of superconductivity in these systems, carefully constructed $T-x$ phase diagrams are needed.
Elemental analysis, preferably of single crystal samples, should be used to determine the actual percentage of the
dopant inside the lattice rather than the nominal doping level. Recently such a detailed study was made for Co doped
${\rm BaFe}_2{\rm As}_2$ \cite{NiCo,chu, nmr, wen, Dan, Lester},
as well as for Ni, Cu, and Cu/Co mixes \cite{threedoping}. These data on $3d$, electron doped ${\rm BaFe}_2{\rm As}_2$ raised
the question of whether the number of impurities, the band filling, and / or the unit cell dimensions were the physically salient variables.
For this paper, ${\rm Ba(Fe}_{1-x}{\rm TM}_x)_2{\rm As}_2$ (TM =
Rh, Pd) series have been studied by the electrical transport, magnetization, specific heat and wave-length dispersive
spectroscopy. We find that the $T-x$ phase diagrams for Co- and Rh-doping are virtually identical,
as are the phase diagrams for Ni- and Pd-doping. By analysis of the relative changes in the unit cell parameters
we can conclude that whereas $x$ and $e$ can still successfully be used to parameterize the structual / magentic
and superconducting phase transitions in the ${\rm Ba(Fe}_{1-x}{\rm TM}_x)_2{\rm As}_2$ systems, changes
in the unit cell parameters, or their ratios, no longer can.

\section{experimental methods}
Single crystals of ${\rm Ba(Fe}_{1-x}{\rm TM}_x)_2{\rm As}_2$(TM =
Rh, Pd) were grown out of self flux using
conventional high-temperature solution growth techniques \cite{NiCo, canfield, threedoping}.
FeAs, RhAs and PdAs powder were synthesized in the same manner as in \cite{NiCo}. Small Ba chunks, FeAs/RhAs or FeAs/PdAs
 powder were mixed together according to the ratio Ba:TMAs = 1:4.
 The mixture was placed into an alumina crucible with a second
 "catch" crucible containing quartz wool placed on top. Both crucibles were
 sealed in a quartz tube under a 1/3, partial atmosphere, of Ar gas. The sealed quartz tube was heated up to 1180 $^\circ $C over 12 hours,
 held at 1180 $^\circ $C for 5 hours, and then cooled to 1050 $^\circ $C over 36 hours. Once the furnace reached
 1050 $^\circ $C, the excess FeAs/RhAs or FeAs/PdAs liquid was decanted from the plate like single crystals.

Powder x-ray diffraction measurements, with a Si standard, were performed using a Rigaku Miniflex diffractometer
with Cu $K_{\alpha}$ radiation at room temperature. Diffraction patterns were taken on ground single crystals from each batch.
No detectable impurities were found in these compounds.
The unit cell parameters were refined by "UnitCell" software. Error bars were taken as twice the standard deviation, $\sigma$,
which was obtained from the refinements by the "Unitcell" software.
Elemental analysis of the samples was performed using wavelength
dispersive x-ray spectroscopy (WDS) in the electron probe microanalyzer
of a JEOL JXA-8200 electron-microprobe.
Magnetization and temperature-dependent AC electrical resistance data (f=16Hz, I=3mA) were collected in a
Quantum Design (QD) Magnetic Properties Measurement System (MPMS) using LR700 resistance bridge for the latter. Electrical contact was made to the sample using Epotek
H20E silver epoxy to attach Pt wires in a four-probe configuration. Heat capacity data were collected using
a QD Physical Properties Measurement System (PPMS) using the relaxation technique.

\section{results}

Summaries of the WDS measurement data are shown in Table I for both
${\rm Ba(Fe}_{1-x}{\rm Rh}_x)_2{\rm As}_2$ and ${\rm Ba(Fe}_{1-x}{\rm Pd}_x)_2{\rm As}_2$.
For each batch, up to 5 pieces of samples were measured. The table shows the number of locations measured,
the average of the $x$ values measured at these locations,
and two times the standard deviation of the $x$ values measured on these locations, which is taken as the error bar in this paper.
We can see that the $2\sigma$ error bars are $\lesssim$ 10\% of the average $x$ values.
The average $x$ value, $x_{ave}$, obtained from wavelength dispersive
x-ray spectroscopy (WDS) measurement will be used for all the compounds in this paper
rather than nominal $x$. It is worth noting that separate measurements of $x_{ave}$
on the resistivity bars gave values within the $2\sigma$ error bars for all the measured batches.

\bet
\begin{tabular}{c|c|c|c|c|c|c|c|c}
   \hline
   \hline
   \multicolumn{9}{c}{${\rm Ba(Fe}_{1-x}{\rm Rh}_x)_2{\rm As}_2$}\\
   \hline
   N & 16 & 16 & 18 & 15 & 20 & 34 & 33 & 20 \\
   \hline
   $x_{ave}$ & 0.012 & 0.026 & 0.039 & 0.057 & 0.076 & 0.096 & 0.131 & 0.171 \\
   \hline
   $2\sigma$ & 0.001 & 0.001 & 0.002 & 0.003 & 0.004 & 0.006 & 0.005 & 0.002 \\
   \hline
   \multicolumn{9}{c}{${\rm Ba(Fe}_{1-x}{\rm Pd}_x)_2{\rm As}_2$}\\
   \hline
   N & 18 & 8 & 52 & 6 & 6 & 12 & 14 & 52 \\
   \hline
   $x_{ave}$ & 0.012 & 0.021 & 0.027 & 0.030 & 0.043 & 0.053 & 0.067 & 0.077 \\
   \hline
   $2\sigma$ & 0.001 & 0.002 & 0.003 & 0.002 & 0.001 & 0.002 & 0.002 & 0.005 \\
   \hline
   \hline
   \end{tabular}
\caption{The WDS data for ${\rm Ba(Fe}_{1-x}{\rm Rh}_x)_2{\rm As}_2$ and ${\rm Ba(Fe}_{1-x}{\rm Pd}_x)_2{\rm As}_2$.
N is the number of locations measured in one batch,
$x_{ave}$ is the average $x$ value measured in one batch,
$2\sigma$ is two times the standard deviation of the N values measured.}
\eet

Fig. \ref{rh} presents the normalized electrical resistivity data of the ${\rm Ba(Fe}_{1-x}{\rm Rh}_x)_2{\rm As}_2$ series
from base temperature, 2 K, to 300 K.
Normalized resistivity, instead of resistivity, is plotted because of the tendancy of these samples
to exfoliate or partially crack \cite{NiCo, tanatar1, tanatar2}. The resistive anomaly at 134 K for pure ${\rm BaFe}_2{\rm As}_2$ is
associated with the structural/magnetic phase transitions \cite{ba122}.
As in the case of Co, Ni and Cu substitutions \cite{NiCo, threedoping}, as $x$ is increased the temperature of the resistive anomaly is
suppressed monotonically
and the shape of the feature changes from a sharp decrease in pure ${\rm BaFe}_2{\rm As}_2$ to a broadened increase in
doped samples. It is no longer detectable for $x\geq0.057$. For ${x=0.026}$, superconductivity becomes detectable,
with ${T_c}$ ${\approx3}$ K inferred from the sharp drop in the resistivity data. For ${x=0.057}$, superconducting temperature $T_c$ has a maximum value
of 24 K with a width ${\Delta T_c}$ ${\approx0.7}$ K. With even higher $x$, $T_c$ is suppressed.

\bef
\psfig{file=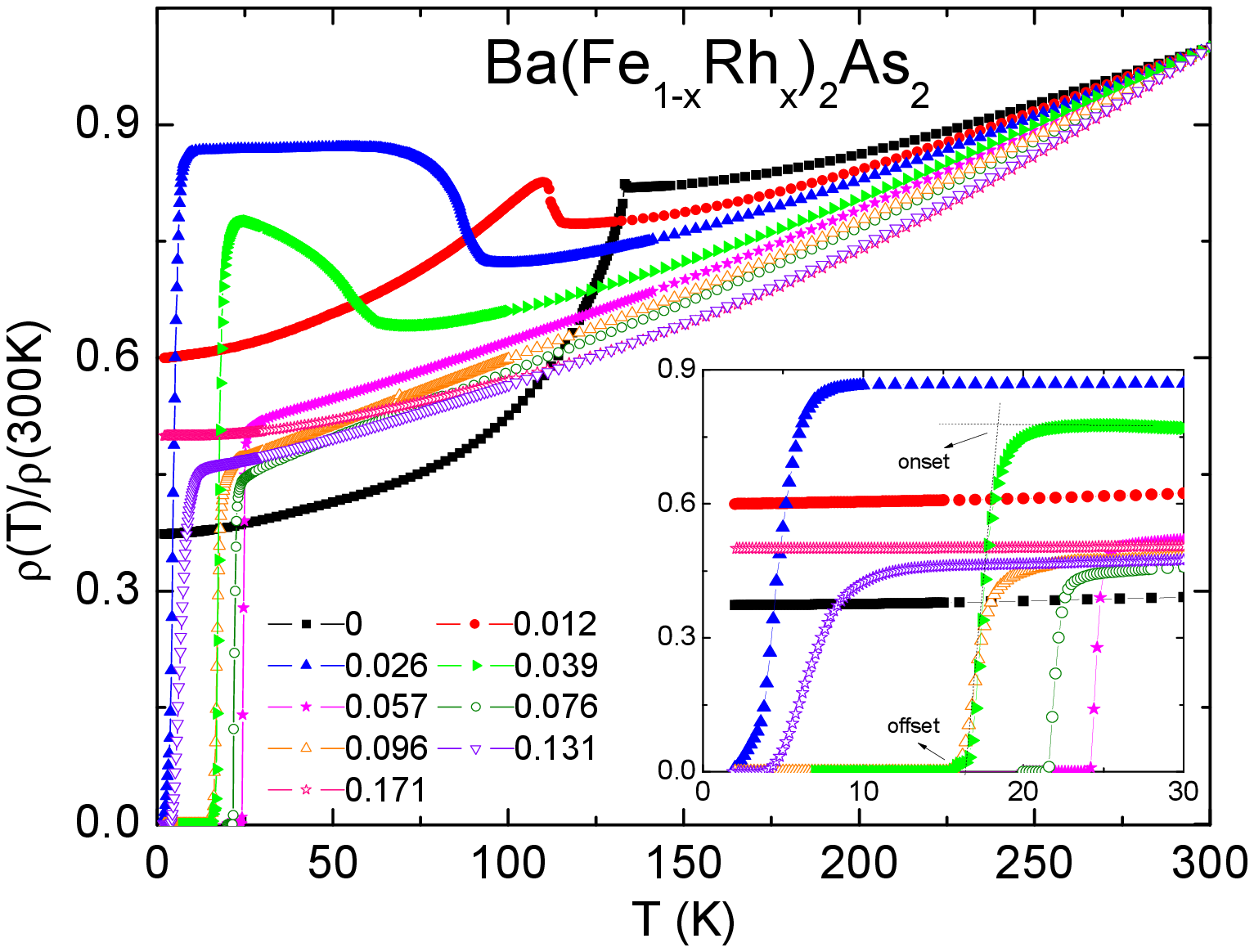,width=3.2in}
\caption{The temperature
dependent resistivity, normalized by the room temperature value, for
${\rm Ba(Fe}_{1-x}{\rm Rh}_x)_2{\rm As}_2$.
Inset: low temperature data for ${\rm Ba(Fe}_{1-x}{\rm
Rh}_x)_2{\rm As}_2$. Onset and offset criteria for $T_c$ are shown.}
\label{rh}
\eef

\bef
\psfig{file=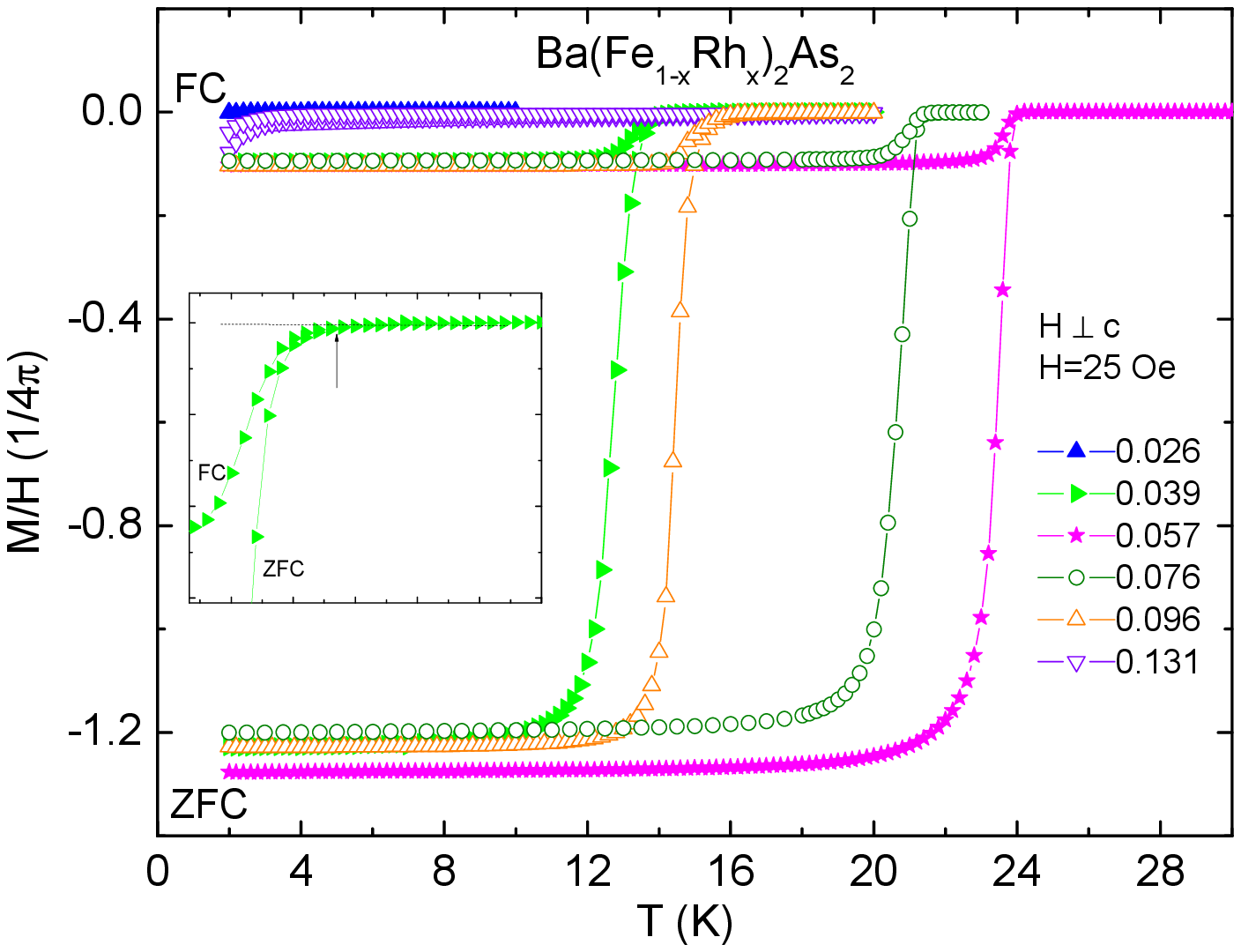,width=3.2in}
\caption{Low magnetic field M/H for
${\rm Ba(Fe}_{1-x}{\rm Rh}_x)_2{\rm As}_2$ series. Inset: The criterion used to infer $T_c$ is shown for ${\rm Ba(Fe}_{0.961}{\rm Rh}_{0.039})_2{\rm As}_2$.
}
\label{mrh}
\eef

Fig. \ref{mrh} shows the $M/H$ data for the ${\rm Ba(Fe}_{1-x}{\rm Rh}_x)_2{\rm As}_2$ series taken at 25 Oe with $H$ perpendicular to the
crystallographic $c$-axis. A clear diamagnetic signal can be seen in both field-cooled(FC) and zero-field-cooled(ZFC) data.
Because of the low $T_c$ values
for ${x=0.026}$ and ${x=0.131}$, which are on the low- and high- $x$ extremes of the superconductivity dome respectively,
we only observe the onset of the diamagnetic signal and no large drop below the superconducting temperature is seen down to our
base temperature of 2 K. However, for all the other concentrations, the large superconducting, shielding fraction
and the sharp drop below $T_c$ are consistent with the existence of bulk superconductivity.
Compared to the low field $M/H$ data for ${\rm Ba(Fe}_{1-x}{\rm Co}_x)_2{\rm As}_2$ \cite{NiCo}, the
superconducting fraction associated with the ${\rm Ba(Fe}_{1-x}{\rm Rh}_x)_2{\rm As}_2$ series have very similar values as of
${\rm Ba(Fe}_{1-x}{\rm Co}_x)_2{\rm As}_2$ series.

The temperature dependent heat capacity data for ${\rm Ba(Fe}_{0.943}{\rm Rh}_{0.057})_2{\rm As}_2$ is shown in Fig. \ref{crh}. This concentration has
the maximum $T_c$ value in this series. The heat capacity
anomaly is relatively sharp and consistent with the superconducting phase transition we observed in both resistivity and low field magnetization data.
The large arrow in the inset shows
the onset of superconductivity and $T_c$ = 23.2 K. A way to estimate $\triangle C_p$ is also shown in the inset; $\triangle C_p\approx$ 700 mJ/mole K.
Assuming the BCS weak coupling approximation $\triangle C_p/\gamma T_c=$ 1.43 and 100\% superconducting volume,
the $\gamma$ value for ${\rm Ba(Fe}_{0.943}{\rm Rh}_{0.057})_2{\rm As}_2$
can be estimated to be about 20 mJ/mole K$^{2}$, which is comparable to the value estimated in the same manner for
${\rm Ba(Fe}_{0.943}{\rm Co}_{0.074})_2{\rm As}_2$ \cite{sergey}.

\bef
\psfig{file=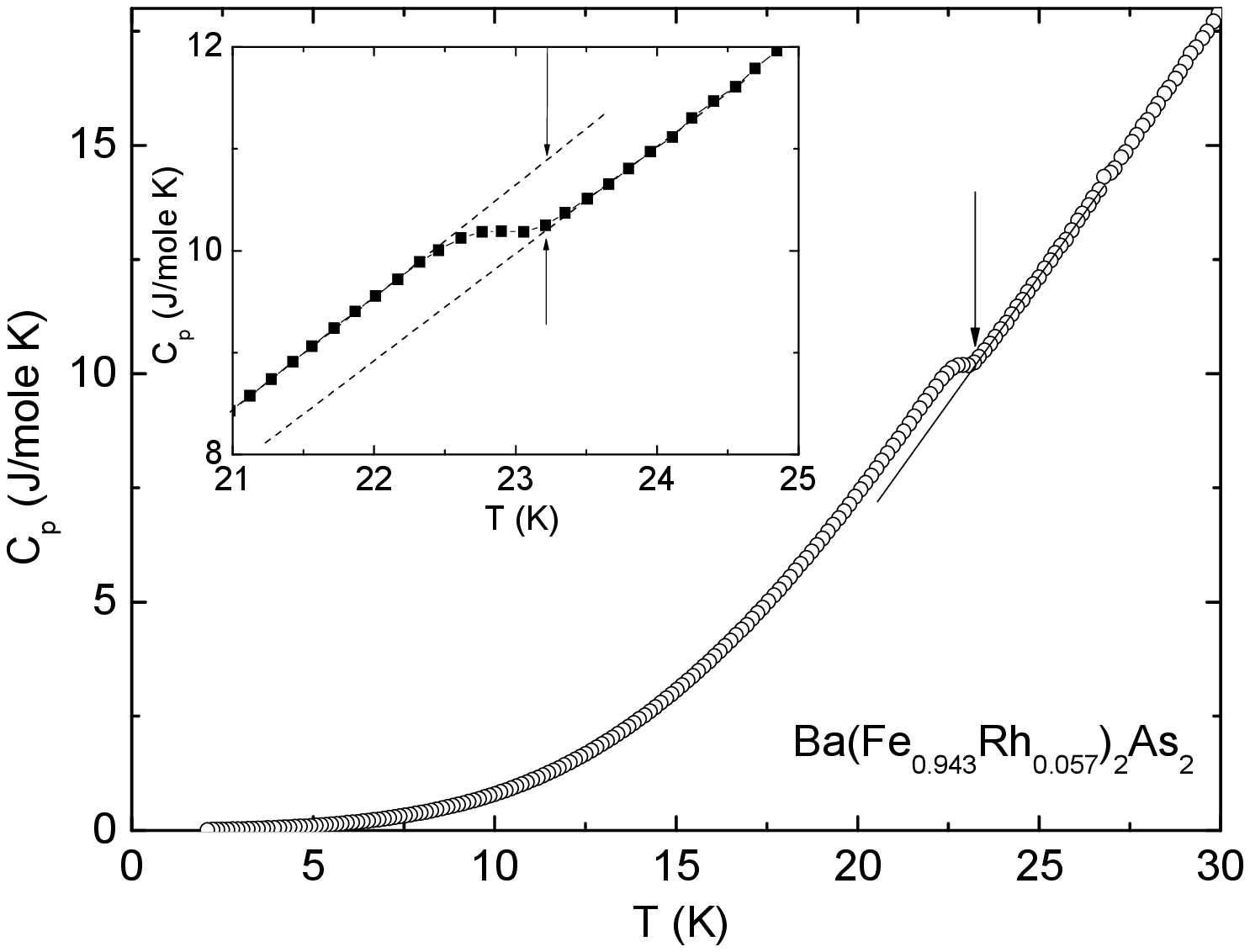,width=3.2in}
\caption{Temperature dependent heat capacity data for ${\rm Ba(Fe}_{0.943}{\rm Rh}_{0.057})_2{\rm As}_2$.
Inset: $C_p$ vs. T near the superconducting transition with the estimated $\triangle C_p$ shown.}
\label{crh}
\eef

\bef
\psfig{file=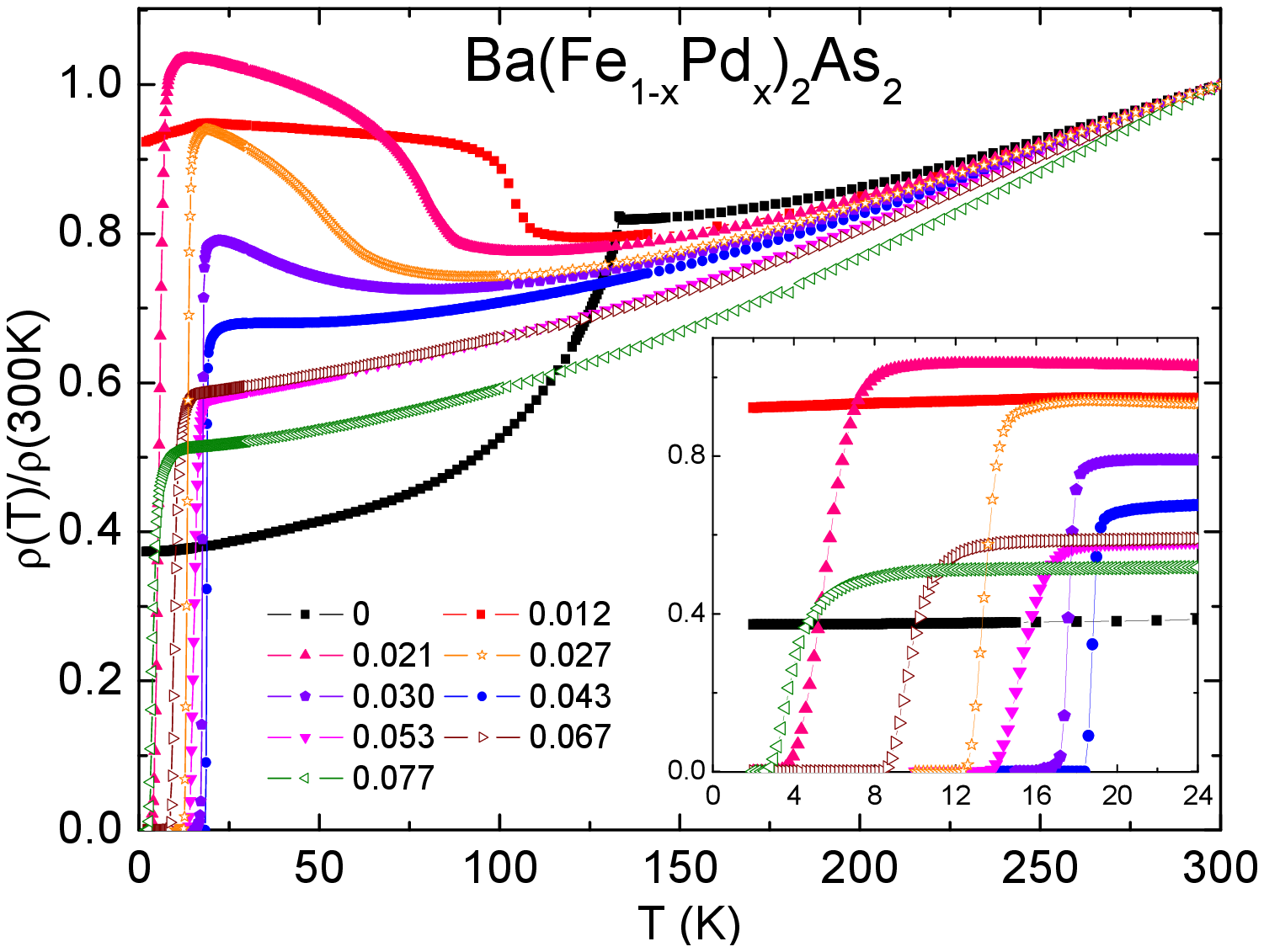,width=3.4in}
\caption{The temperature
dependent resistivity, normalized by room temperature value, for
${\rm Ba(Fe}_{1-x}{\rm Pd}_x)_2{\rm As}_2$.
Inset: low temperature data for ${\rm Ba(Fe}_{1-x}{\rm
Pd}_x)_2{\rm As}_2$}
\label{pd}
\eef

Fig. \ref{pd} shows the normalized electrical resistivity data for the ${\rm Ba(Fe}_{1-x}{\rm Pd}_x)_2{\rm As}_2$ series from base temperature, 2 K,
to 300 K.
A systematic behavior, similar to the ${\rm Ba(Fe}_{1-x}{\rm Rh}_x)_2{\rm As}_2$ series, is seen:
the temperature of the resistive anomaly associated with the structural / antiferromagnetic phase transitions
is suppressed monotonically with Pd doping and the shape of the anomaly changes from a sharp decrease
to a broadened increase in resistivity upon cooling. For $x=0.021$, the resistive anomaly can still be clearly seen and superconductivity
is detected with ${T_c}$ ${\approx5.7}$ K. For $x=0.043$, the temperature of the resistive anomaly is further reduced and it is only inferred
from a minimum in the resistivity above the superconducting transition. For $x=0.053$, the resistive anomaly is
completely suppressed and $T_c$ has its highest value of about 19 K and a width of ${\Delta T_c}$ ${\approx0.6}$ K.
With higher $x$ values, $T_c$ is suppressed.

\bef
\psfig{file=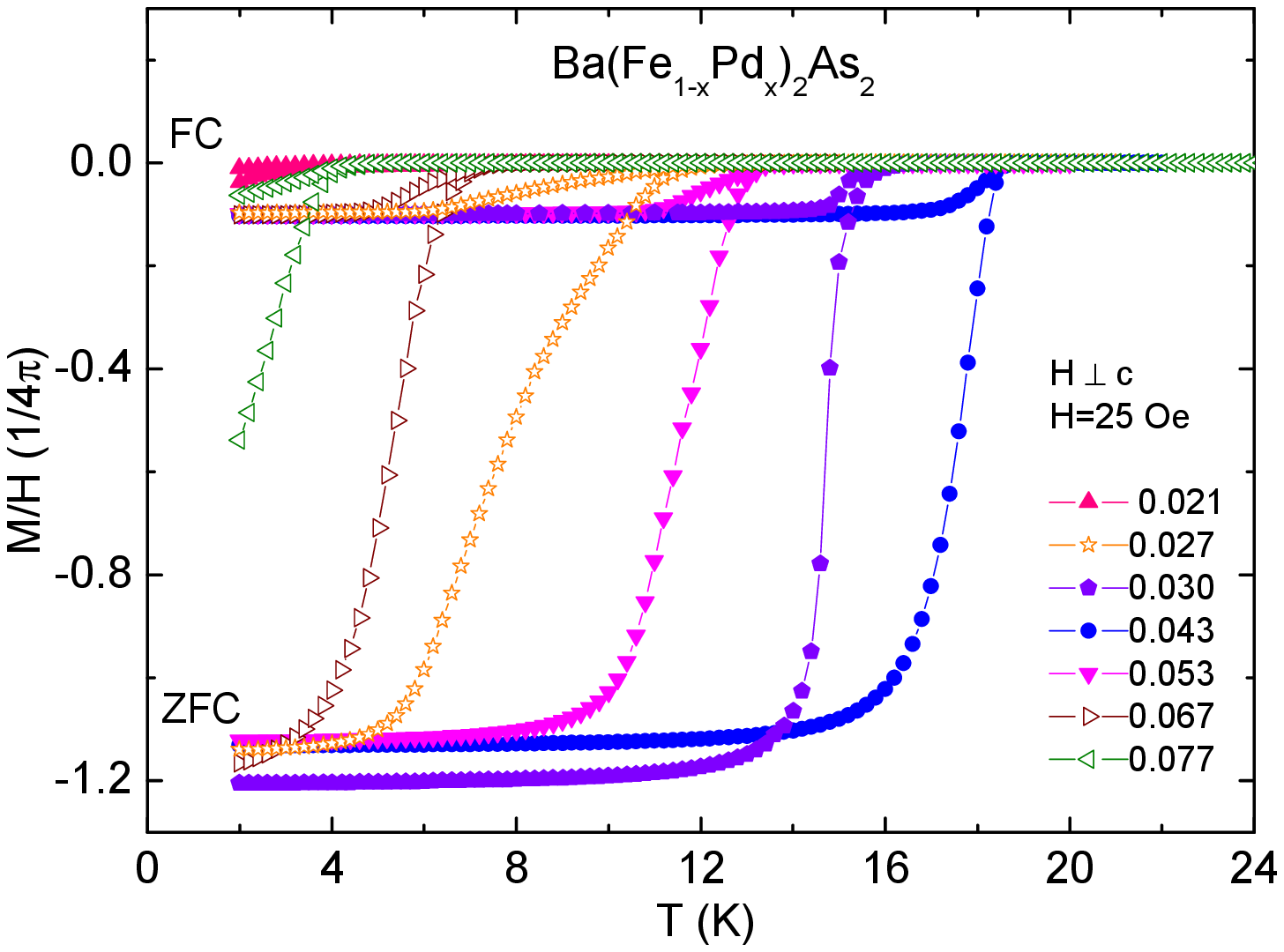,width=3.2in}
\caption{Low magnetic field $M/H$ for
${\rm Ba(Fe}_{1-x}{\rm Rh}_x)_2{\rm As}_2$ series.
}
\label{mpd}
\eef

The low field $M/H$ data for the ${\rm Ba(Fe}_{1-x}{\rm Pd}_x)_2{\rm As}_2$ series (FC and ZFC) are shown in Fig. \ref{mpd}.
They were taken at 25 Oe with $H$ perpendicular to the crystallographic $c$-axis.
The broader feature seen in the magnetization for $x=0.027$ implies a larger inhomogeneity associated with this sample. Indeed, the WDS
data for $x=0.027$ does show local maximum in $2\sigma$ values.
Despite the broader drop of the magnetization, the large superconducting fraction is comparable
to the rest of the Pd- doped series as well as to the Co-, Ni- and Rh- doped ${\rm BaFe}_2{\rm As}_2$ results, all of which are consistent
with bulk superconductivity.
Again, only a small diamagnetic signal was observed at base temperature for $x=0.021$ due to the low $T_c$ for this concentration.

Fig. \ref{cpd} shows the temperature dependent heat capacity data for ${\rm Ba(Fe}_{0.957}{\rm Pd}_{0.043})_2{\rm As}_2$, which manifests the
highest $T_c$ value in this series. The heat capacity
anomaly at $T_c$ can be clearly seen, although it is broader than the one found for ${\rm Ba(Fe}_{0.943}{\rm Rh}_{0.057})_2{\rm As}_2$ (Fig. \ref{crh}).
The arrows show the onset of superconductivity at $T_c=18$ K, and the estimated $\triangle C_p$ is shown in the inset; $\triangle C_p\approx$ 410 mJ/mole K.
Using the BCS weak coupling approximation $\triangle C_p/\gamma T_c$ = 1.43 and assuming 100\% superconducting volume in this sample,
$\gamma$ for ${\rm Ba(Fe}_{0.957}{\rm Pd}_{0.043})_2{\rm As}_2$ is estimated to be about 16 mJ/mole K$^{2}$.

\bef
\psfig{file=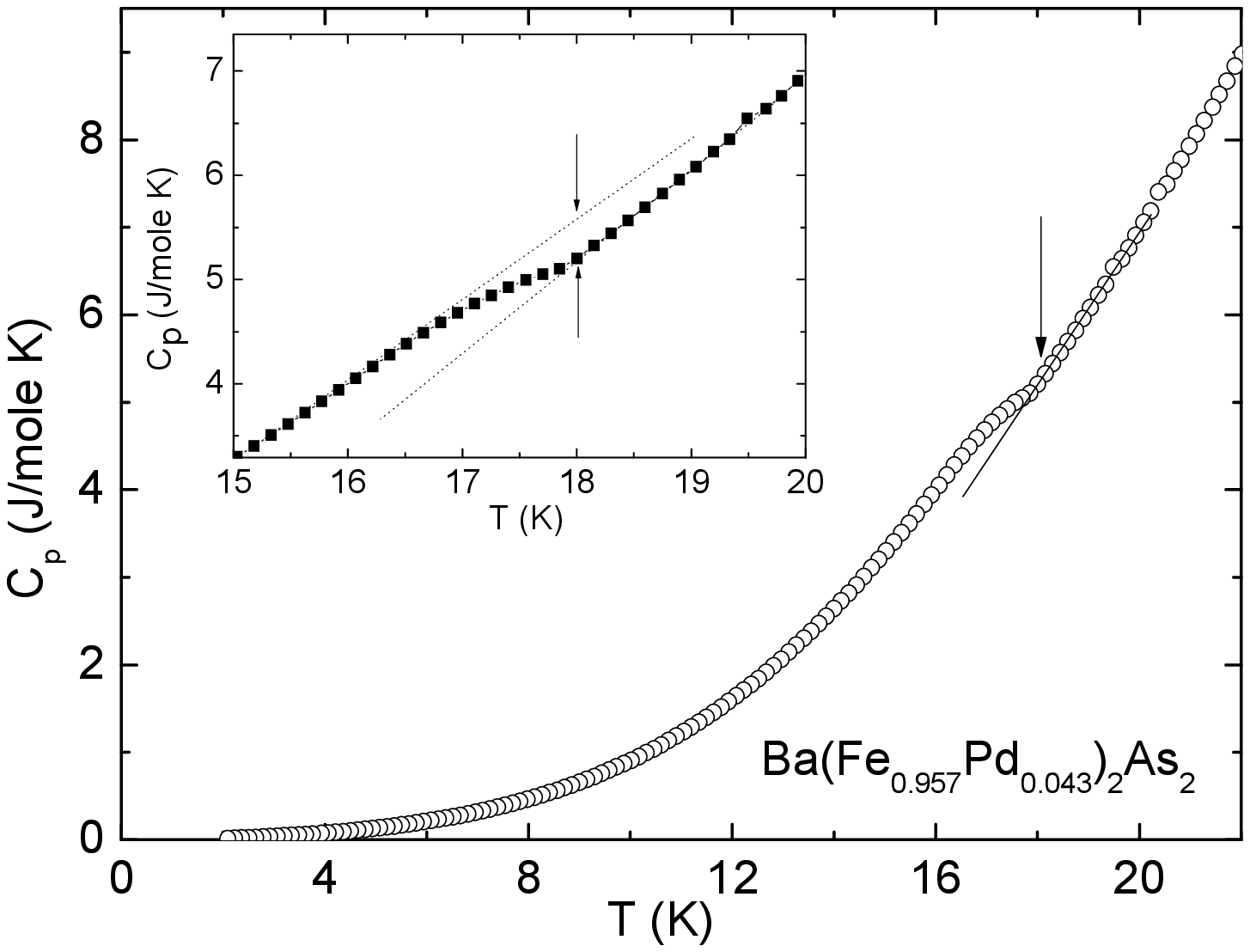,width=3.2in}
\caption{Temperature dependent heat capacity data for ${\rm Ba(Fe}_{0.957}{\rm Pd}_{0.043})_2{\rm As}_2$.
Inset: $C_p$ vs. T near the superconducting transition with the estimated $\triangle C_p$ shown.}
\label{cpd}
\eef

\section{discussion}

The data presented in Figs. 1 - 6 are summarized in the two, ${T-x}$, phase
diagrams shown in Fig. \ref{tx}. In this paper, the temperature of structrural/magnetic phase transitions
are inferred from the derivative of the temperature dependent resistivity data which shows a split feature for finite values of $x$ \cite{NiCo}.
Onset and offset criteria, which are shown in the inset
of Fig. \ref{rh}, are used to determine $T_c$ from the resistivity data. The criterion which is shown in the inset of
Fig. \ref{mrh} is used to determine $T_c$ from the magnetization data. The arrows in Fig. \ref{crh} show the criterion used to infer
$T_c$ from heat capacity data. We can see good agreement between resistivity, magnetization and heat capacity measurements.
So as to allow comparison with the isoelectronic, $3d$ electron doped ${\rm BaFe}_2{\rm As}_2$ compounds,
data for ${\rm Ba(Fe}_{1-x}{\rm TM}_x)_2{\rm As}_2$ (TM = Co, Ni) \cite{NiCo, threedoping} are also shown.

\bef
\psfig{file=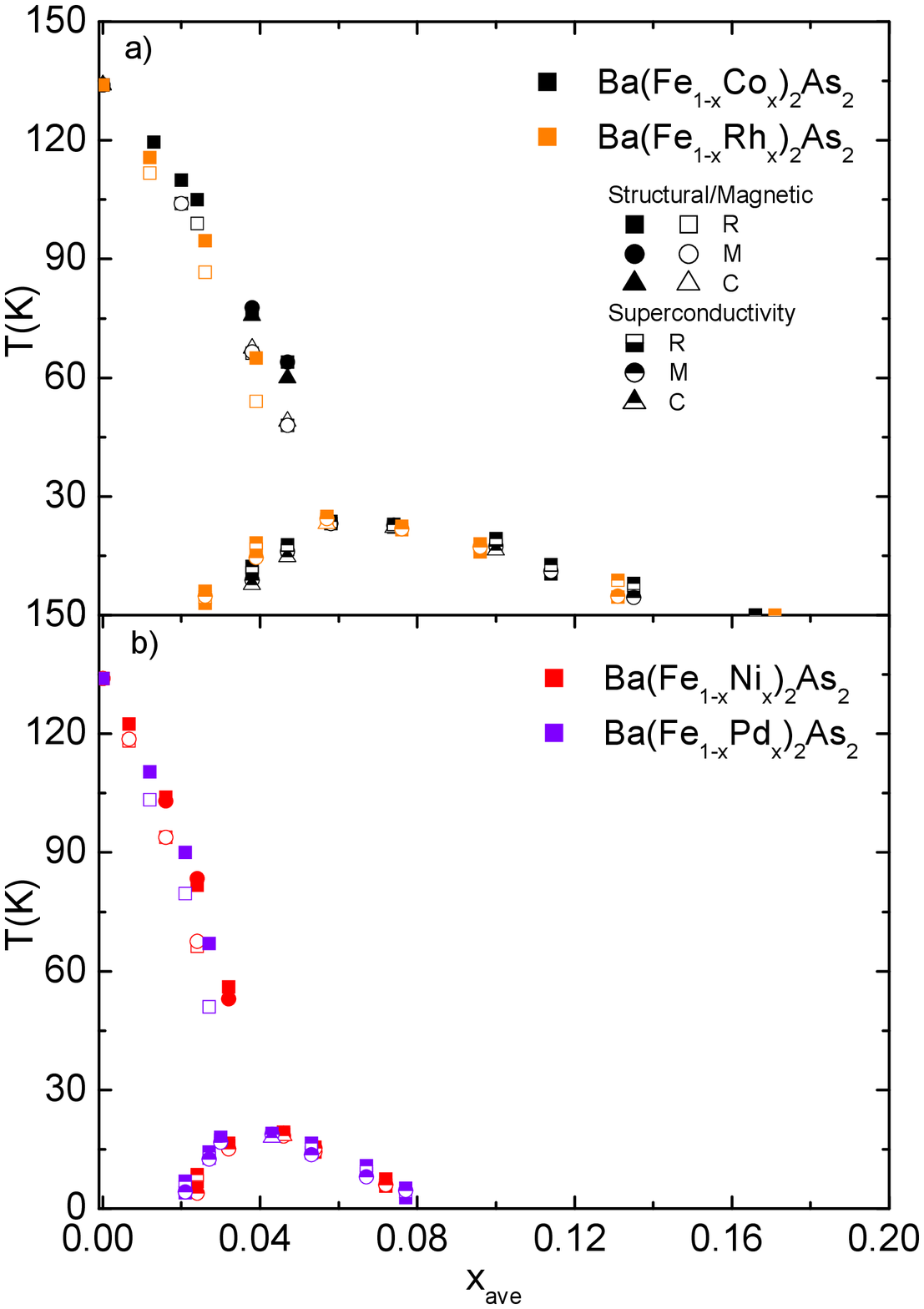,width=3.4in}
\caption{ Transition temperature as a
function of $x$. (a): $T-x$ phase diagrams of ${\rm Ba(Fe}_{1-x}{\rm Rh}_x)_2{\rm As}_2$ and ${\rm Ba(Fe}_{1-x}{\rm Co}_x)_2{\rm As}_2$ series.
(b): $T-x$ phase diagrams of ${\rm Ba(Fe}_{1-x}{\rm Pd}_x)_2{\rm As}_2$ and ${\rm Ba(Fe}_{1-x}{\rm Ni}_x)_2{\rm As}_2$ series. For both plots the transition temperatures were determined
in a manner similar to that described in \cite{NiCo} and the text.}
\label{tx}
\eef

The upper panel presents the $T-x$ phase diagrams of Rh and Co doped ${\rm BaFe}_2{\rm As}_2$ and the lower panel
presents the $T-x$ phase diagrams of Pd and Ni doped ${\rm BaFe}_2{\rm As}_2$.
It can be seen in both panels that the higher temperature structural/magnetic phase transitions are suppressed monotonically in a similar manner/rate
for all series. Superconductivity is found in both tetragonal and orthorhombic phase \cite{NiCo,chu, nmr, wen}, and is stabilized in
a dome-like region for all series. Superconductivity is found over a
wider range of Co or Rh doping with a maximum ${T_c}$ around 24 K, and a narrower range of
Ni or Pd doping with a maximum $T_c$ around 19 K.

The complete phase diagram of Rh doped ${\rm BaFe}_2{\rm As}_2$
including both structural/magnetic phase transition and superconductivity shows incredible similarity as the phase diagram of
Co doped ${\rm BaFe}_2{\rm As}_2$ and the complete phase diagram of Pd doped ${\rm BaFe}_2{\rm As}_2$
shows incredible similarity as the phase diagram of Ni doped ${\rm BaFe}_2{\rm As}_2$.
For each of the pairs, the phase diagrams show exceptionally similar behavior on the rate of the suppression of structural/magnetic
phase transitions, the range of superconducting domes and the maximum $T_c$. It is worth mentioning again that
to see this, the actual $x$ values are vital.

\bef \psfig{file=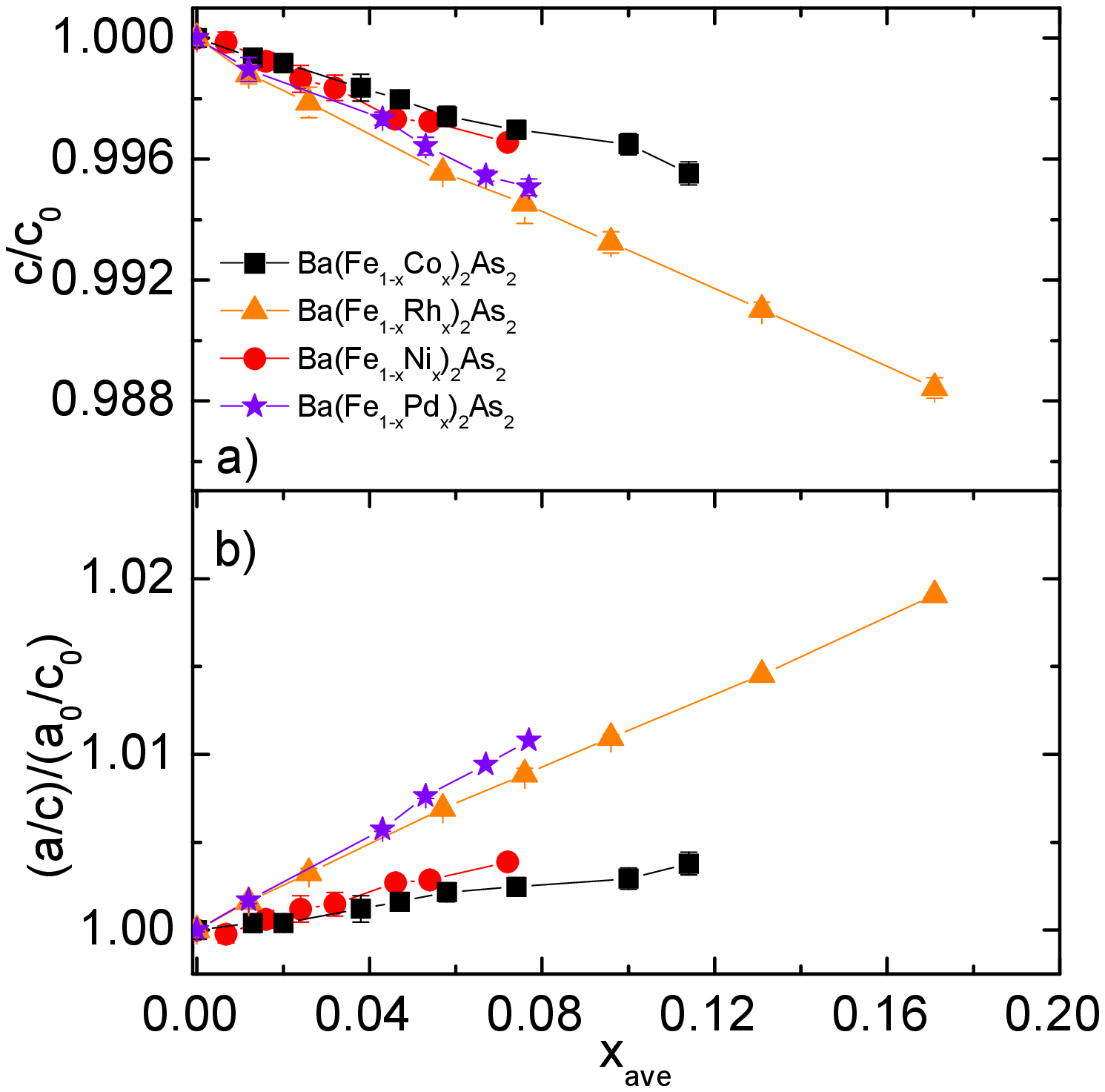,width=3.2in} \caption{Normalized
structural parameters measured at $\sim$ 300 K. (a) $a/a_0$, (b) $(a/c)/(a_0/c_0)$ as a function of transition metal doping,
$x$. ($a_0$=3.9621(4) $\AA$, $c_0$=13.0178(10) $\AA$)} \label{lattice} \eef

In our previous work \cite{threedoping}, we compared the
transition temperatures as a function of $x$, and as a function of the number of extra conduction electrons, $e$,
added by the dopant per Fe/TM site for Co, Ni, Cu and Co/Cu doped ${\rm BaFe}_2{\rm As}_2$ (for the case of Co ${e = x}$, for the
case of Ni ${e = 2x}$, for the case of Cu ${e = 3x}$). We conclude that whereas the suppression of the
structural/antiferromagnetic transitions was parameterized by the
number of TM dopant ions (or, equivalently, changes in the $c$-axis) the
superconducting dome was parameterized by
the number of electrons added by doping (or, equivalently, changes in the values of the ${a/c}$
ratio), and exists over a limited range of $e$-values (or band filling).
Unfortunately we could not experimentally separate the effects of $x$ and $e$ from changes in $c$ and $a/c$, respectively \cite{threedoping}.
However, with current, $4d$ electron doped ${\rm BaFe}_2{\rm As}_2$ data, we can actually distinguish between $x$ and $e$ on one hand and $c$, $a/c$ on the other.
Fig. \ref{lattice} shows the unit cell parameters
normalized by the lattice parameters of pure ${\rm BaFe}_2{\rm As}_2$. To compare the unit cell parameters, the data for Co or Ni doped
${\rm BaFe}_2{\rm As}_2$ \cite{NiCo, threedoping} are also plotted in Fig.\ref{lattice}. The lattice parameter $c$ decreases with all dopings and the ratio of
$a/c$ increases with all dopings. But in both cases there is a clear difference between the $3d$- and $4d$- data sets.
Unlike in our previous work \cite{threedoping}, where for $3d$ electron doping series and $c$ can be scaled with
$x$, $a/c$ can be scaled with the number of extra
electron added per Fe/TM site, when $4d$ electron doped ${\rm BaFe}_2{\rm As}_2$ data are taken into account, changes in
$c$ and $a/c$ are no longer equivalent to $x$ and $e$. This means that if we want to parameterize the effects of $3d$- and $4d$-TM doping
on the transitions temperatures of
${\rm Ba(Fe}_{1-x}{\rm TM}_x)_2{\rm As}_2$, whereas the upper, structural and magnetic phase transitions
can be parameterized by $x$ and the superconducting dome can be parameterized by $e$, they are no longer well parameterized
by either $c$ or $a/c$.  As discussed in \cite{threedoping}, it is still possible that some other parameter, such as bonding angles
associated with the As position, offer better or alternate parameterization of these transition temperatures, but these are not currently known.

\bef
\psfig{file=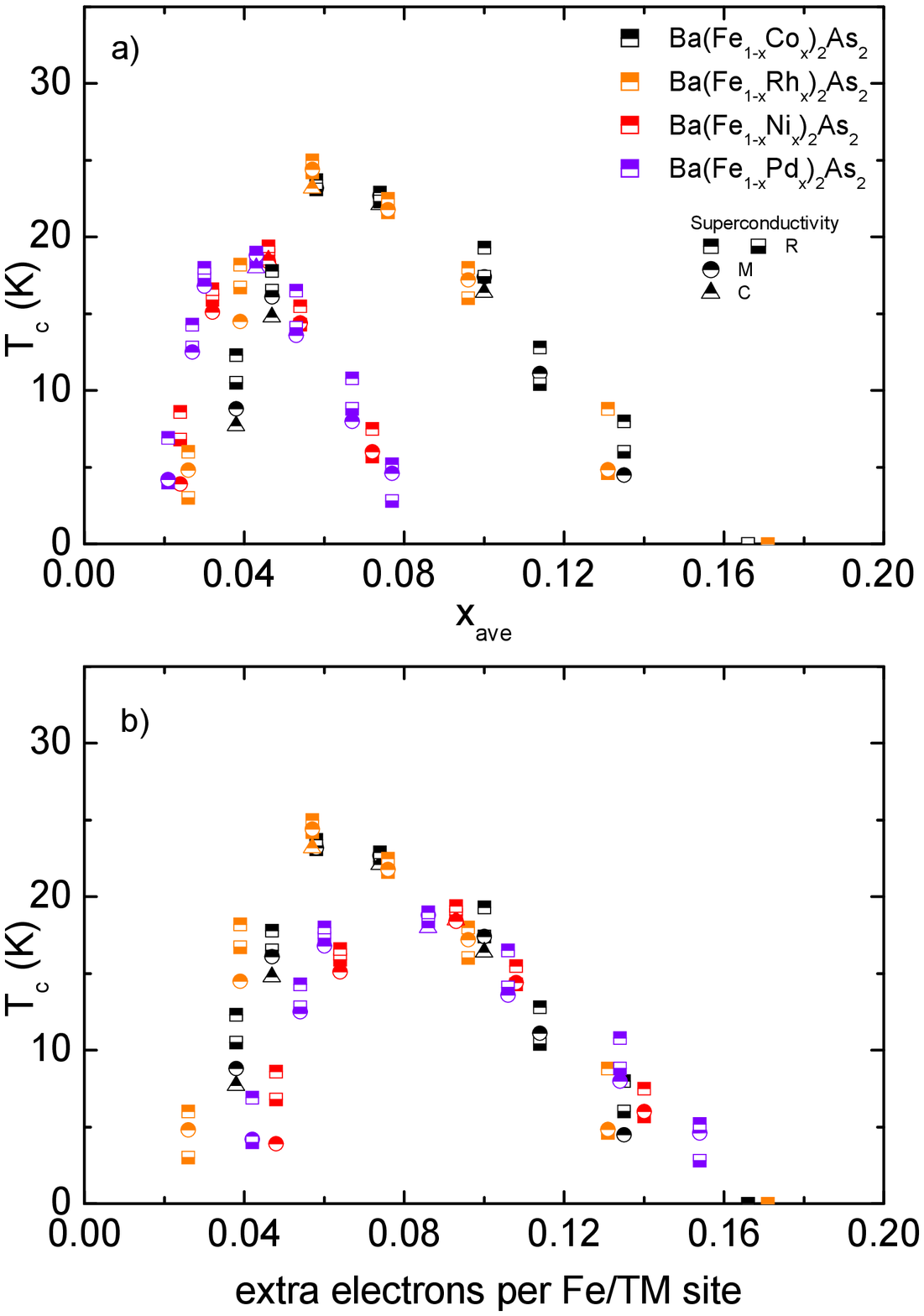,width=3.4in}
\caption{(a): superconducting transition temperature $T_c$ as a
function of $x_{ave}$. (b): superconducting transition temperature $T_c$ as a
function of extra electrons per Fe/TM site.
}
\label{tetx}
\eef

The parameterization scheme outlined above is based on the premise that a single parameter may be controlling the variation
of the upper, structural and magnetic transition temperatures ($T_{s/m}$) and a second one may be controlling the superconductivity. There is another scheme that should be
discussed in the context of our growing data set: that there may be a single parameter that controls the behavior of the system
when $T_{s/m}$ $>$ $T_c$ and there is another single parameter when $T_{s/m}$ is fully suppressed. The potential appeal of this scheme can be seen in Fig. \ref{tetx},
where $T_c$ is plotted as a function of $e$ and $x$ for comparison. As discussed above and in reference \cite{threedoping},
there is excellent agreement of the $T_c$ values
when plotted as a function of $e$ when $T_{s/m}$ is fully suppressed. On the other hand there is arguably better agreement of the $T_c$ values when they
are plotted as a function of $x$ for $T_{s/m}$ $>$ $T_c$.  As pointed out in ref \cite{threedoping} the behavior on the $T_{s/m}$ $>$ $T_c$ side of the
dome may be associated with the need to bring the upper transition to low enough temperature to allow the superconductivity to turn on.
The importance of reducing $T_{s/m}$ may be associated with reducing the degree of orthorhombic splitting, the size of the ordered moment
in the AF phase, and / or changing the magnetic excitation spectrum.

\bef
\psfig{file=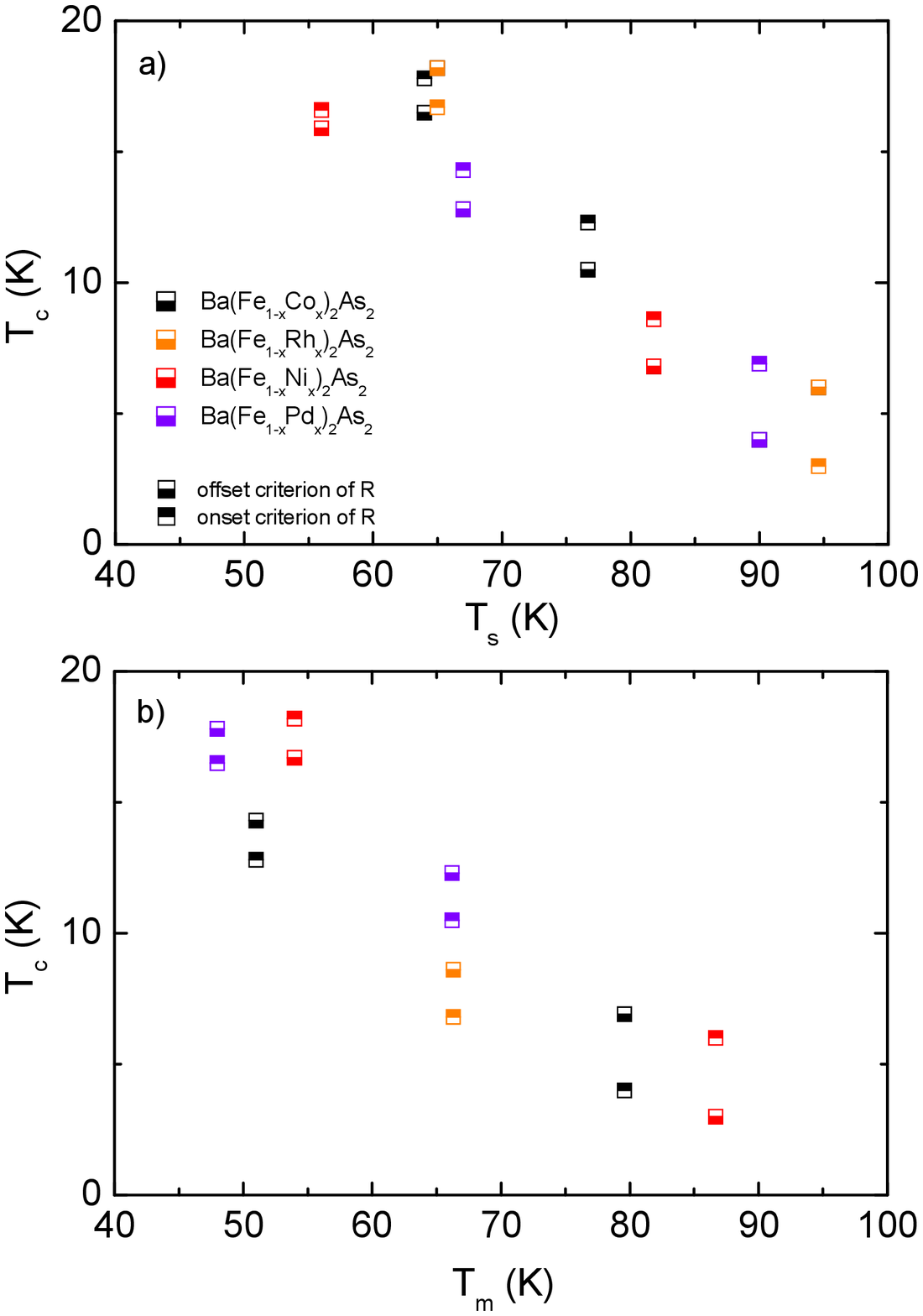,width=3.4in}
\caption{(a): Superconducting transition temperature $T_c$ as a
function of structural phase transition temperature $T_s$. (b): Superconducting transition temperature $T_c$ as a
function of magnetic phase transition temperature $T_m$.
}
\label{tstmtc}
\eef

Given that $T_{s/m}$ only roughly scales with $x$ it is worth while examining the correlation between $T_s$, $T_m$ and $T_c$ more directly.
Fig. \ref{tstmtc} plots $T_c$ as a function of the structural, as well as the magnetic, transition temperature (given that they are
split by the time superconductivity is stabilized) \cite{NiCo, Lester, Dan}. Both plots show a clear correlation. A more graphic way of examining
the correlation between $T_c$ and $T_{s/m}$ is to create a composite diagram for the $T_{s/m}$ $>$ $T_c$ data by adjusting the $x$ scales for the
Ni, Rh, and Pd data so as to collapse the $T_s$ and $T_m$ phase lines onto the Co data set. This is plotted in Fig. \ref{xmix}.
As we can see, a clear consequence of this is to bring collapse the $T_c$ data onto a single phase line as well.

\bef
\psfig{file=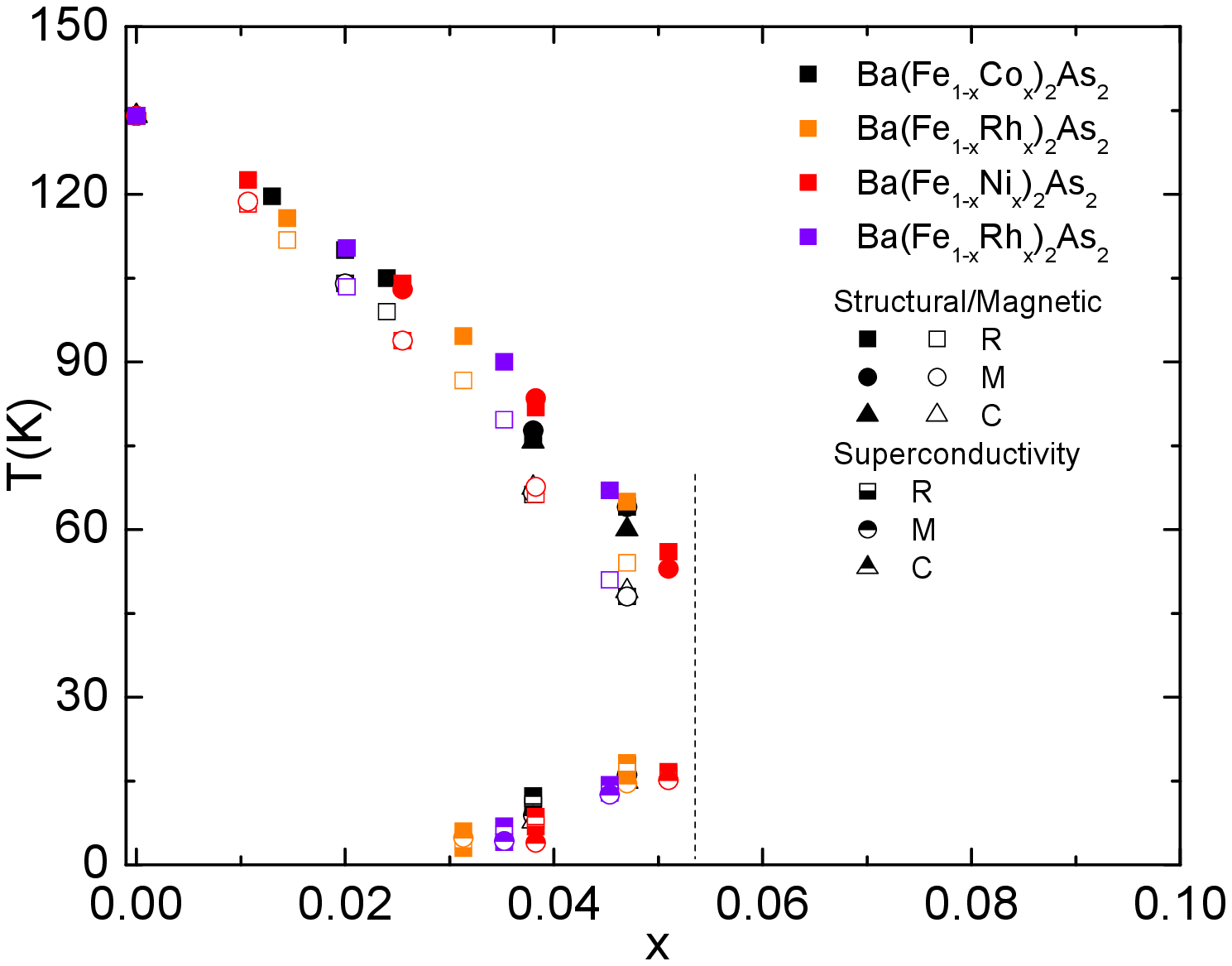,width=3.4in}
\caption{ Transition temperature as a
function of adjusted $x$. $x$ is normalized so as to bring the interpolated values of $T_s$ onto the transition associated with
${\rm Ba(Fe}_{0.953}{\rm Co}_{0.047})_2{\rm As}_2$: for
Co doped ${\rm BaFe}_2{\rm As}_2$, $x=x_{ave}$; for
Rh doped ${\rm BaFe}_2{\rm As}_2$, $x=x_{ave}\times0.047/0.039$; for Pd doped ${\rm BaFe}_2{\rm As}_2$, $x=x_{ave}\times0.047/0.028$;
for Ni doped ${\rm BaFe}_2{\rm As}_2$, $x=x_{ave}\times0.047/0.03$.
}
\label{xmix}
\eef

\section{conclusion}

Single crystalline ${\rm Ba(Fe}_{1-x}{\rm TM}_x)_2{\rm As}_2$ (TM = Rh, Pd) samples have been grown and characterized by microscopic,
thermodynamic and transport measurements.
$T-x$ phase diagrams were constructed for both the Rh- and Pd-doping series and, remarkably, they are virtually indistinguishable
from the $T-x$ phase diagrams assembled for their $3d$-shell counterpart, Co- and Ni-doped, series. Given that the variations of
the unit cell parameters are distinctly different for the $3d$ and $4d$ dopants, these data clearly show that whereas the amount
of dopant, $x$, and the change in electron count, $e$, do a fair job of parameterizing the structural / magnetic and
superconducting phase transitions temperatures, respectively, the variation of the c-axis lattice parameter and
the variation of the ratio of the $a/c$ parameters no longer do.

Whereas the structural and magnetic phase transitions are fairly well parameterized by $x$ and, for $T_c$ $>$ $T_{s/m}$, $T_c$ is parameterized
by $e$ very well, the $T_c$ data for $T_{s/m}$ $>$ $T_c$ appears to depend on the degree of suppression of $T_{s/m}$ (and therefore may depend more on
$x$ than on $e$).  The fact that the behavior of $T_c$ in response to doping appears to change in the vicinity of the disappearance of
$T_{s/m}$ is consistent with recent studies of the $T-P$ phase diagram for ${\rm BaFe}_2{\rm As}_2$ \cite{estelle} as well as earlier work on
K-doped ${\rm BaFe}_2{\rm As}_2$ \cite{kphase}.
In every case $T_c$ appears to reach its maximum value (varying from TM-, to K-, to P-doped \cite{P}) when $T_{s/m}$ is suppressed below $T_c$.

\section*{Acknowledgments}
Work at the Ames Laboratory was supported by
the Department of Energy, Basic Energy Sciences under Contract No.
DE-AC02-07CH11358. We would like to thank M. Tanatar, C. Martin, E. Colombier, E. D. Mun, M. E. Tillman, S. Kim, X. Lin for help and useful discussions.


\end{document}